\documentclass{emulateapj}

\shorttitle{OBSERVATIONS OF H$_{2}$ EMISSION FROM AB AURIGAE}
\shortauthors{BITNER ET AL.}

\newcommand{\szero}{S(0)}
\newcommand{\sone}{S(1)}
\newcommand{\stwo}{S(2)}
\newcommand{\sfour}{S(4)}

\begin{document}

\title{TEXES OBSERVATIONS OF PURE ROTATIONAL H$_{2}$ EMISSION FROM AB AURIGAE}

\author{Martin A. Bitner\altaffilmark{1,2}, Matthew J. Richter\altaffilmark{2,3}, 
John H. Lacy\altaffilmark{1,2}, Thomas K. Greathouse\altaffilmark{2,4}, 
Daniel T. Jaffe\altaffilmark{1,2}, Geoffrey A. Blake\altaffilmark{5}}

\altaffiltext{1}{Department of Astronomy, University of Texas at Austin, Austin, TX 78712; 
mbitner@astro.as.utexas.edu, lacy@astro.as.utexas.edu, dtj@astro.as.utexas.edu}
\altaffiltext{2}{Visiting Astronomer at the Infrared Telescope Facility, which is operated by the University of Hawaii under Cooperative Agreement no. NCC 5-538 with the National Aeronautics and Space Administration, Science Mission Directorate, Planetary Astronomy Program.}
\altaffiltext{3}{Physics Department, University of California at Davis, Davis, CA 95616; 
richter@physics.davis.edu}
\altaffiltext{4}{Lunar and Planetary Institute, Houston, TX 77058; greathouse@lpi.usra.edu}
\altaffiltext{5}{Division of Geological \& Planetary Sciences, California Institute 
of Technology, MS 150-21, Pasadena, CA 91125; gab@gps.caltech.edu}

\begin{abstract}
We present observations of pure rotational molecular hydrogen 
emission from the Herbig Ae star, AB Aurigae.
Our observations were made using the Texas Echelon Cross Echelle 
Spectrograph (TEXES) at the NASA Infrared Telescope Facility and the Gemini 
North Observatory.  
We searched for H$_{2}$ emission in the \sone, \stwo, and \sfour ~lines at 
high spectral resolution and detected all three.
By fitting a simple model for the emission in the three 
transitions, we derive T $= 670 \pm 40$ K and 
M $=0.52 \pm 0.15$~M$_{\oplus}$ for the emitting gas.
Based on the 8.5 km s$^{-1}$ FWHM of the \stwo ~line, assuming the emission 
comes from the circumstellar disk, and with an inclination estimate of the AB Aur 
system taken from the literature, we place the location for the emission near 18 AU.
Comparison of our derived temperature to a disk structure model suggests 
that UV and X-ray heating are important in heating the disk atmosphere.
\end{abstract}




\section{INTRODUCTION}

Disks around young stars are a natural part of the star 
formation process and, as the likely site of planet 
formation, have generated much interest.  
A considerable 
amount has been learned about the structure of circumstellar 
disks through the study of dust emission and detailed modeling 
of the spectral energy distributions (SED) of these stars.  
However, if we assume the standard gas-to-dust ratio 
for the interstellar medium, dust makes up only 1\% of the 
mass of the disk.  
Thus there is an interest in developing direct tracers of the gas component.
Studies of gas in disks have so far focused mainly on gas 
at either large radii ($> 50$ AU), using observations at 
submillimeter wavelengths \citep{semenov05}, or small radii, 
using observations of near-infrared CO lines \citep{blake04,najita03}.  
Mid-infrared spectral diagnostics may probe disks at 
intermediate radii (1-10 AU) \citep{najita06}.

Molecular hydrogen diagnostics are promising because they trace the dominant 
constituent of the disk and so do not rely on conversion factors 
to determine the mass of the emitting gas.  
H$_{2}$ has been observed in circumstellar environments at ultraviolet 
\citep{krull00} and near-infrared \citep{bary03} wavelengths.  
These observations trace hot circumstellar gas, or gas excited 
by fluorescent processes, and are therefore difficult to translate into gas 
masses.  
The pure rotational mid-infrared H$_{2}$ lines are useful probes 
because the level populations should be in LTE at the local gas 
temperature, and so line ratios allow determination of the 
excitation temperature and mass of the warm gas.  
The line ratios of the three low-J lines accessible from the ground are 
sensitive to gas at temperatures of 200-800 K.
Other lines in the mid-infrared such as [NeII] at 12.8 $\mu$m 
\citep{glassgold07}, [SI] at 25.2 $\mu$m and [FeII] at 26 $\mu$m 
\citep{pascucci06} should be useful probes of 
gas in disks but their interpretation requires detailed modeling.
Molecular hydrogen emission does not trace the cold, optically thick regions 
of disks, and therefore estimates of the amount of warm gas based on 
H$_{2}$ emission represent a lower limit to the total disk mass. 
In order to see H$_{2}$ emission arising from a disk, the warm gas must be 
either physically separated from optically thick dust or at a 
different temperature.  
These conditions can be met in the disk surface layer or in gaps or holes 
in the disk.
In either case, gas can be physically separated (due to dust settling or clearing) 
as well as thermally decoupled (due to lower densities) from the 
dust and heated to a higher temperature by X-ray and UV radiation 
\citep{glassgold04,gorti04,nomura05}.

From the ground, three H$_{2}$ mid-infrared rotational lines are accessible:  
\sone ~($\lambda = 17.035 ~\mu$m), \stwo ~($\lambda = 12.279 ~\mu$m), and
\sfour ~($\lambda = 8.025 ~\mu$m).   
The high spectral resolution achievable with the Texas Echelon Cross 
Echelle Spectrograph (TEXES; Lacy et al. 2002) is crucial for 
these observations in order to separate the lines from nearby telluric 
features and to maximize the line contrast against the dust continuum.  
Observations at high spectral resolution also have the  
advantage that, when coupled with information about the disk inclination, 
they allow an estimate for the location of the emitting gas.  
For emission lines with small equivalent width, 
ground-based observations with TEXES can be more sensitive than 
\textit{Spitzer} IRS observations.

AB Aur is one of the brightest and thus one of the most well-studied of 
the Herbig Ae stars.
It is located at a distance of 144 pc based on Hipparcos measurements 
\citep{vanden98}, 
has a spectral type of A0-A1 \citep{hernandez04} and is surrounded 
by a disk/envelope structure that extends to at least $r \sim 450$ AU \citep{mannings97}.  
Observations at 11.7 and 18.7 $\mu$m by \citet{chen03} show that AB Aur 
varies in the mid-infrared.  
The mass of the star is 2.4 M$_\odot$ with an age of 2-4 Myr \citep{vanden98}.  
AB Aur has a SED which is well-fit by the passive irradiated disk with puffed-up inner rim model of \citet{dullemond01}.  
There have been a wide range of inclination estimates for AB Aur (see 
Brittain et al. 2003 for an extended discussion of this issue).  
Recent results by \citet{semenov05} derive an inclination of $17 ^{+6}_{-3}$ deg 
for the AB Aur system by modeling millimeter observations.  We will assume this 
value for analysis of our data.  

Several attempts have been made to observe molecular hydrogen emission from 
the disk of AB Aur.
\citet{thi01} claimed the detection of H$_{2}$ \szero ~and \sone ~emission associated with 
the AB Aur system using the \textit{Infrared Space Observatory (ISO)}, but 
subsequent observations from the ground with improved spatial resolution 
were unable to confirm the emission \citep{richter02}.  
\citet{richter02} and \citet{sheret03} saw some evidence of emission in the \stwo 
~line but the data quality was such that a definitive statement could not 
be made.
They saw no evidence of the \sone ~line.  
In this paper, we present definite detections of the \sone, \stwo, and \sfour ~lines. 


\section{OBSERVATIONS AND DATA REDUCTION}

All of the data were acquired using TEXES in its high-resolution mode on 
the NASA Infrared Telescope Facility (IRTF) between December 2002 
and October 2004 and on Gemini North in November 2006 under program ID 
GN-2006B-Q-42.  
Table~1 lists details of the observations.
We nodded the source along the slit to remove background sky emission.
We observed flux and telluric standards at each setting.  
Other calibration files taken include blank sky and an ambient temperature 
blackbody used to calibrate the wavelength and to flatfield the data.  

The data were reduced using the standard TEXES pipeline \citep{lacy02} which produces 
wavelength-calibrated one dimensional spectra.  
The TEXES pipeline data reduction gives a first order flux calibration.  
However, because of the different slit illumination of the ambient temperature 
blackbody and point sources, we observed flux standards to determine a correction factor.
We assumed that the slit illumination was the same for the target as for the flux 
standard.  
Guiding and seeing variations can introduce uncertainty into our flux calibration.


\section{RESULTS}
\begin{figure}
\plotone{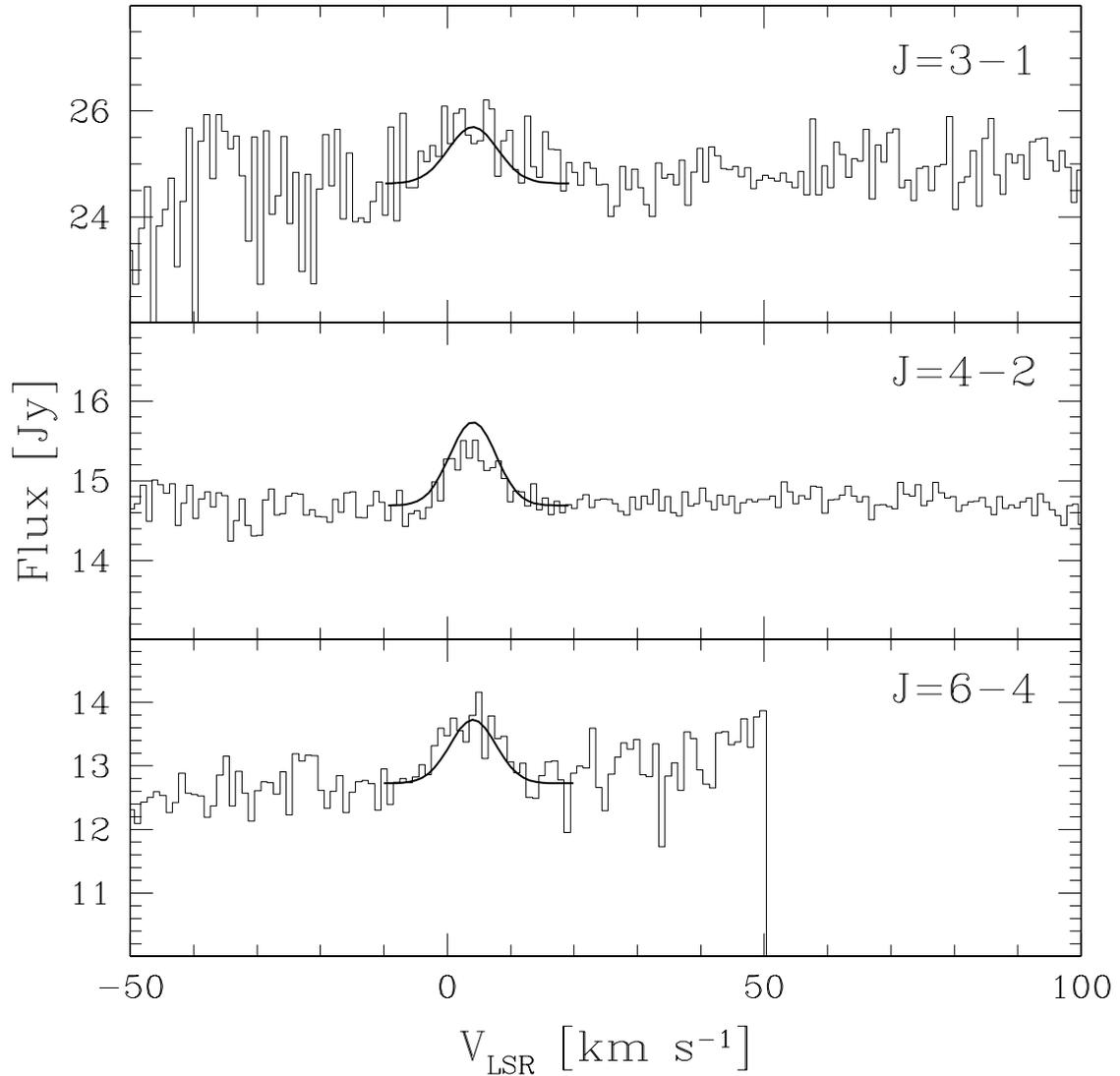}
\caption{Regions of the three H$_{2}$ pure rotational lines available from the ground.  
The data shown were obtained at Gemini North in November 2006 under Gemini 
program GN-2006B-Q-42. 
Each spectrum has been flux-calibrated.  
The lines were individually fit with a Gaussian to determine their centroid, 
flux, and FWHM. 
The lines are all centered near the systemic velocity of AB Aur and have FWHM 
$\sim$10 km s$^{-1}$.
Line fluxes are listed in Table~2.
The overplotted Gaussians are based on the simultaneous fit and are not the individual fits 
to each line.
The three spectral regions were simultaneously fit with a model assuming the 
observed emission arises from a mass of H$_{2}$ gas at a single temperature.
The best-fit model with  T = 670 K and H$_{2}$ mass = 0.52 M$_{\oplus}$ is overplotted.  
\label{3linefit-fig}
}
\end{figure}

Figure~\ref{3linefit-fig} shows flux-calibrated spectra for all three 
settings taken at the Gemini North Observatory.  
Overplotted on each spectrum is a Gaussian with integrated line flux 
equal to that derived from the best-fit model assuming the emission 
arises from an isothermal mass of optically thin H$_{2}$ gas.  
The three lines were fit simultaneously while varying the temperature 
and mass of the emitting gas in the model.
For the simultaneous fitting, we adopted the FWHM and centroid of the \stwo ~line for all three lines.
The quoted errors for the temperature and mass are 1-$\sigma$ based on the contour plot of 
the $\chi^{2}$ values.  
Our best-fit model based on Gemini data has T~$=670 \pm 40$ K and M~$=0.52 \pm0.15 $~M$_{\oplus}$.   
A similar analysis of our IRTF data, which consisted of a detection at \stwo ~and 
upper limits for \sone ~and \sfour, gives T~$=630 \pm 60$ K and M~$=0.7 \pm0.2 $~M$_{\oplus}$.
Though within 2-$\sigma$ uncertainties, the lower measured \stwo ~flux observed at Gemini than from the 
IRTF combined with the fact that our slit is smaller on the sky at Gemini suggests we may be 
resolving out some of the flux we see at the IRTF.
To investigate this possibility, we examined spatial plots of our 2-D echellograms but found 
no clear evidence for the existence of spatially resolved line emission.

Table~2 contains a summary of our results.
We fit each line individually with a Gaussian to determine centroid, FWHM, and flux values.  
The lines are all centered near the systemic velocity of AB Aur and have FWHM $\sim$10 km s$^{-1}$.
\citet{roberge01} find A$_{v}$ = 0.25 for AB Aur, so we assume no extinction at our wavelengths.  
We quote equivalent widths in addition to line fluxes due 
to uncertainty in the determination of the continuum level.
In an attempt to test how sensitive our temperature and mass estimates are to errors 
in the flux calibration, we fit our data after normalizing to the continuum flux values 
from ISO SWS observations of AB Aur \citep{meeus01}.
The resulting temperature and mass do not differ significantly from the values based on our 
internal flux calibration.
We derive T $=630$ K and M $=0.71$~M$_{\oplus}$ when normalizing to the ISO continuum values of 13.1 Jy 
at 8 $\mu$m, 21.0 Jy at 12 $\mu$m, and 30.8 Jy at 17 $\mu$m before fitting.
The upper limits for \sone ~and \sfour ~from the IRTF quoted in Table~2 are based on 
a Gaussian fit at the expected position of each line plus a 1-$\sigma$ error.  
The 1-$\sigma$ line flux errors were computed by summing over the number 
of pixels corresponding to the FWHM of the lines in regions of the spectrum 
with similar atmospheric transmission.   

Figure~\ref{popdia-fig} shows a population diagram based on our observations.  
The points marked are based on the line fluxes derived from Gemini observations 
listed in Table~2, and the error bars are 1-$\sigma$.
The overplotted solid line is not a fit to the population diagram; rather it 
is based on the best-fit model temperature and mass.  
Even though we only have three points, the deviation from a single temperature 
(a straight line on this plot) is significant and may indicate that the emission 
is coming from a mix of temperatures, as expected for a disk with a radial temperature 
gradient.
We note that curvature in the population diagram can also occur if the ortho-to-para (OTP) 
ratio is different from our assumed value of 3.
\citet{fuente99} derived a an OTP ratio between 1.5-2 for H$_{2}$ in a photodissociation 
region with gas temperatures of 300-700 K.  
However, the sense of the curvature in our population diagram would require an 
OTP ratio greater than 3 to explain the increased \sone ~flux, which seems unlikely.
We discuss the derived temperature of the gas and its location in the next section.  


\section{DISCUSSION}

We have detected all three mid-IR H$_{2}$ lines with good signal to noise.
The results are consistent with the previous upper limits and the 2-$\sigma$ detection 
of the \stwo ~line \citep{richter02,sheret03}.
Our observations of all three lines allow us to put tighter 
constraints on the temperature and mass of the emitting gas.
The analysis of our data has three components.  
First, we determine a characteristic temperature of the emitting gas by fitting a simple model 
to the three lines.
Second, we locate the position in the disk where the emission arises by comparing our observed 
line profiles to line profiles calculated from a Keplerian disk model.
Finally, in order to comment on the structure of the disk, we compare our results to the predictions 
of a well-established disk structure model which assumes well-mixed gas and dust at the same 
temperature and find that, under such assumptions, we do not reproduce our observations.  
This argues for the need for additional gas heating mechanisms which may come in the form 
of X-ray and UV heating.  The importance of UV and X-ray heating in the surface layers of 
disks has been demonstrated observationally by \citet{bary03} and \citet{qi06}.

\begin{figure}
\includegraphics[scale=.80,angle=90]{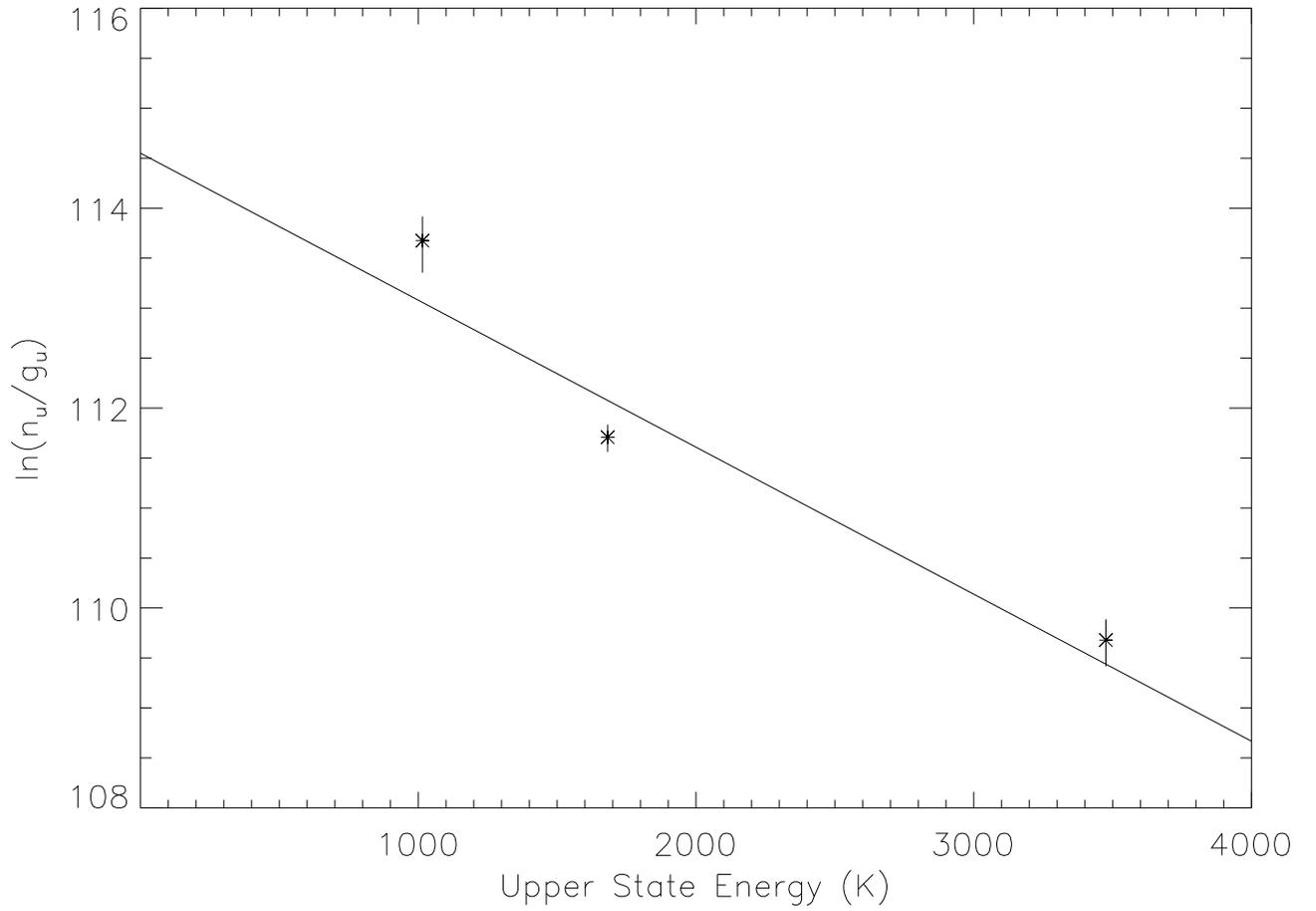}
\caption
{Population diagram based on Gemini North observations.
The points marked are based on fluxes derived by fitting a Gaussian to each of the lines and 
shown with 1-$\sigma$ error bars.
The overplotted solid line is not a fit to the population diagram; rather it is 
based on the best-fit model temperature and mass found by simultaneously fitting 
all three lines.
The deviation of the points from the single-temperature line is significant and 
suggests that the emission may arise from gas at a range of temperatures.
\label{popdia-fig}
}
\end{figure}

The procedure used to derive a temperature of T~$=670 \pm 40$ K and mass of 
M~$=0.52 \pm0.15$~M$_{\oplus}$ from a simulaneous fit to the H$_{2}$ \sone, \stwo, and \sfour 
~lines assumes the gas is optically thin and in LTE at a single temperature.
If the emission arises in the atmosphere of a disk where radial and vertical 
temperature gradients exist, the derived temperature 
should be considered a characteristic or average temperature. 
In fact, Figure~\ref{popdia-fig} shows evidence that the emission arises in 
gas at a mix of temperatures.
It should be noted that radial temperature gradients in a disk also affect the 
detectability of various lines.
Since we are most sensitive to narrow emission lines, the \sone ~line, originating in 
cooler gas farther out in the disk, is easier to detect than emission in the \sfour ~line.
Emission arising very close to the star from the inner rim of the disk would be 
more difficult for us to detect at our high spectral resolution due to line broadening.

Our derived temperature, 670 K, is in agreement with the 
temperature found by \citet{richter02} who derived T $> 380$ K based on \stwo ~and 
the upper limit on \sone.
\citet{brittain03} made observations toward AB Aur at 4.7 $\mu$m of 
$^{12}$CO v = 1-0 emission.
The low-J lines in the resulting population diagram have a steeper slope, 
interpreted as emission coming from cooler gas than the high-J lines.
Fits to the population diagrams assuming LTE gas give T $= 70$ K for the cool gas 
and T $= 1540$ K for the hot gas.  
The authors explain the hot CO emission as coming from the inner rim of the disk, 
while the cool emission originates in the outer flared part of the disk after it 
emerges from the inner rim's shadow.
\citet{blake04} also observed 4.7 $\mu$m CO emission from AB Aur. 
Isothermal fits with T $\sim 800$ K are able to reproduce their line fluxes.
The line widths for the CO lines are generally broader than what we observed, 
suggesting that the CO emission comes from smaller radii or outflowing gas
 \citep{blake04}.

Our derived mass, 0.52 ~M$_{\oplus}$, is four orders of magnitude less than the total 
disk mass of AB Aur \citep{semenov05}.  
This is partly due to the fact that AB Aur's disk is optically thick 
at mid-infrared wavelengths \citep{sheret03}.
Most of the gas in the disk is hidden by optically thick dust and is also too cool 
to emit at our wavelengths.
The mass we derive should therefore be interpreted as a lower limit to the amount 
of warm gas on the surface of the disk facing us.

By comparing the FWHM of our observed \stwo ~line profile with computed line profiles from a simple 
Keplerian disk model, we derive an estimate of where the line emission originates.  
We generate line profiles under the assumption of emission from a Keplerian disk 
inclined at $17^\circ$ \citep{semenov05} orbiting a 2.4 M$_\odot$ star \citep{vanden98} 
with equal emissivity at all points within an annulus 2 AU in extent.
We convolve the line profiles with  our instrumental broadening function plus a thermal 
broadening function based on our derived temperature, T = 670 K, and compare the 
resulting FWHM of the line profiles to our observed \stwo ~line.
Under these assumptions, the line profiles imply that our emission arises near 18 AU in the disk.

We used parameters for the disk based on the fit to AB Aurigae's SED from \citet{dullemond01} 
as input to the vertical disk 
structure model of \citet{dullemond02} in order to derive a grid of temperature 
and density values as a function of disk radius and height above the midplane.
The \citet{dullemond01} model does an excellent job fitting the SED, although we note 
that their derived inclination angle, $65^\circ$, is significantly larger than our assumed 
value of $17^\circ$.
As noted in \citet{dullemond01}, the large discrepancy between their derived 
inclination and what most observations suggest is probably due to the inclusion 
in their SED model of a perfectly vertical inner rim which may not be physically accurate.
We input this radial and vertical temperature and density profile into an LTE radiative 
transfer program.
We assume the gas is in LTE with the dust and compute the emergent 
spectrum.
The emission lines produced by our model under these conditions are much weaker than 
our observations.
This is not especially surprising given our assumption of well-mixed gas and dust at 
the same temperature and suggests the need for additional gas heating mechanisms.

The vertical structure model of \citet{dullemond02} is mainly concerned with modeling the 
dust emission from disks.
By assuming equal gas and dust temperature, the gas only reaches a temperature of $\sim$ 260 K 
at the top of the disk at 18 AU, much cooler than our derived temperature of 670 K. 
To explain the observed emission, an additional mechanism is needed to heat the gas.
X-ray and UV heating are likely possibilities which can heat the gas to temperatures 
significantly hotter than the dust.
\citet{glassgold07} computed the gas temperature in X-ray irradiated disks around T Tauri stars.
At 20 AU, the temperature can reach over 3000 K at the top of the disk before dropping to 500-2000 K 
in a transition zone and then to much cooler temperatures deep in the disk.
\citet{nomura05} considered UV heating of protoplanetary disks and the resulting 
molecular hydrogen emission.
The temperature in their model at 10 AU reaches over 1000 K and predicts an H$_{2}$ \stwo 
~line flux of $2 \times 10^{-15}$ ergs s$^{-1}$ cm$^{-2}$, somewhat lower than our 
observed flux, but their model was for a less massive star and disk.
H$_{2}$ emission could also arise in regions where there is spatial 
separation of the gas from the dust due to dust settling or coagulation of dust 
into larger particles.
The SED of AB Aur shows no evidence for gaps in the AB Aur disk and we are unable 
to make definitive statements based on our observations.

A UV-heated H$_{2}$ layer in a protostellar disk ought to exhibit at least some similarities 
to the comparable hot layer at the surface of photodissociation regions (PDRs) with high 
densities and strong radiation fields.
We therefore compare our observations of AB Aur with earlier mid-IR H$_{2}$ spectroscopy of the 
Orion Bar PDR.
The UV field expressed in terms of the mean interstellar radiation field that was adopted 
by \citet{allers05} ($3 \times 10^{4}$) is somewhat smaller than that expected at 18 AU 
around an A0 star ($\sim 10^{5}$).
\citet{jonkheid07} have computed the strength of the UV field throughout a disk around a 
Herbig Ae star finding values between $10^{5}$ - $10^{6}$ near the disk surface at 20 AU.
Our derived temperature, 670 K, is in the range of temperatures (400-700 K) derived for 
the Orion bar PDR based on the H$_{2}$ \sone, \stwo, and \sfour 
~lines \citep{allers05}.
As shown in \citet{allers05}, the line ratios are determined by a complex mix of the temperature 
gradient, H$_{2}$ abundance and dust properties.
As another check of the plausibility that the surface of AB Aur's disk is similar to a PDR, we computed 
the surface brightnesses of our three lines.
Assuming our observed emission arises from an annulus 2 AU in extent centered at 18 AU, we 
derive surface brightnesses of $\sim 10^{-5}$ ergs  cm$^{-2}$ s$^{-1}$ sr$^{-1}$, roughly an order of 
magnitude lower than those seen by \citep{allers05}.
However, \citet{allers05} estimated that their line intensities were brightened by a 
factor of 10 since the Orion Bar PDR is seen nearly edge-on making our surface brightnesses 
consistent with each other.

\acknowledgements
We thank the Gemini staff, and John White in particular, for their support 
in getting TEXES to work on Gemini North.
We thank Rob Robinson for useful discussions on the analysis of our data.
The development of TEXES was supported by grants from the NSF and the 
NASA/USRA SOFIA project.
Modification of TEXES for use on Gemini was supported by Gemini 
Observatory.  
Observations with TEXES were supported by NSF grant AST-0607312.
This work is based on observations obtained at the Gemini Observatory, 
which is operated by the Association of Universities for Research in 
Astronomy, Inc., under a cooperative agreement with the NSF on behalf 
of the Gemini partnership: the National Science Foundation (United States), 
the Particle Physics and Astronomy Research Council (United Kingdom), 
the National Research Council (Canada), CONICYT (Chile), the Australian 
Research Council (Australia), CNPq (Brazil) and CONICET (Argentina).



\clearpage

\begin{deluxetable}{cccccc}
\tablecaption{Observing Parameters\label{tab:obs}}
\tablehead{
\colhead{Date} &
\colhead{Telescope} &
\colhead{Line\tablenotemark{a}} &
\colhead{Resolving Power\tablenotemark{b}} &
\colhead{Slit Width\tablenotemark{c}} &
\colhead{Integration Time}\\
&
&
&
\colhead{($R\equiv \lambda/ \delta \lambda$)} &
\colhead{(arcsec, AU)} &
\colhead{(s)}
}

\startdata
Dec 2002 & IRTF & \sone & 60,000 & 2.0, 288 & 2591 \\
Dec 2002 & IRTF & \stwo & 85,000 & 1.4, 202 & 4792 \\
Dec 2003 & IRTF & \stwo & 81,000 & 1.4, 202 & 2072 \\
Oct 2004 & IRTF & \sfour & 88,000 & 1.4, 202 & 8549 \\
Nov 2006 & Gemini & \sone & 80,000 & 0.81, 117 & 2007 \\
Nov 2006 & Gemini & \stwo & 100,000 & 0.54, 78 & 2072 \\
Nov 2006 & Gemini & \sfour & 100,000 & 0.54, 78 & 1295 \\ 

\enddata
\tablenotetext{a}{\sone\ is at 587.0324~cm$^{-1}$. \stwo\ is at
814.4246~cm$^{-1}$.  \sfour\ is at 1246.098~cm$^{-1}$. }
\tablenotetext{b}{The resolving power is deduced from
observations of stratospheric emission from Titan or Saturn
at similar frequencies or
unresolved emission lines in a low pressure gas cell.}
\tablenotetext{c}{The slit width in AU is calculated at AB Aur's 
distance of 144 pc.}
\end{deluxetable}

\clearpage

\begin{deluxetable}{cccccc}
\tablecaption{Summary of Results\label{tab:res}}
\tablehead{
\colhead{Telescope} &
\colhead{$\lambda$} &
\colhead{F$_{\nu}$} &
\colhead{Line Flux} &
\colhead{Equivalent Width} &
\colhead{FWHM}\\
&
\colhead{($\mu$m)}&
\colhead{(Jy)} &
\colhead{($10^{-14}$ ~ergs ~s$^{-1}$ ~cm$^{-2}$)} &  
\colhead{(km ~s$^{-1}$)} &
\colhead{(km ~s$^{-1}$)}
}

\startdata
 & 8 & 12.9 (1.1) & $<1.2$ & ... & ... \\
IRTF & 12 & 13.3 (0.4) & 0.93 (0.25) & 0.86 (0.23) & 7.0 \\
 & 17 & 25.2 (1.8) & $<1.1$ & ... & ...  \\
 & 8 & 12.7 (0.4) & 1.47 (0.34) & 0.93 (0.21) & 10.4 \\
Gemini & 12 & 14.7 (0.3) & 0.53 (0.07) & 0.44 (0.06) & 8.5 \\
 & 17 &  24.6 (0.4) & 1.1 (0.30) & 0.76 (0.21) & 9.0 \\

\enddata

\end{deluxetable}

\clearpage

\clearpage

\begin{thebibliography}{}

\bibitem[Allers et al.(2005)]{allers05}
Allers, K.N., Jaffe, D.T., Lacy, J.H., Draine, B.T.,\& Richter, M.J.
2005, \apj, 630, 368

\bibitem[Bary et al.(2003)]{bary03}
Bary, J.S., Weintraub, D.A., \& Kastner, J.H.
2003, \apj, 586, 1136

\bibitem[Beckwith et al.(1990)]{beckwith90}
Beckwith, S.V.W., Sargent, A.I., Chini, R.S., \& Guesten, R.
1990, \aj, 99, 924

\bibitem[Blake \& Boogert(2004)]{blake04}
Blake, G.A., \& Boogert, A.C.A. 
2004, \apj, 606, L73

\bibitem[Brittain et al.(2003)]{brittain03}
Brittain, S.D., Rettig, T.W., Simon, T., Kulesa, C., DiSanti, M.A., 
\& Dello Russo, N.  
2003, \apj, 588, 535

\bibitem[Chen \& Jura(2003)]{chen03}
Chen, C.H., \& Jura, M. 
2003, \apj, 591, 267

\bibitem[Dullemond et al.(2001)]{dullemond01}
Dullemond, C.P., Dominik, D. \& Natta, A.
2001, \apj, 560, 957

\bibitem[Dullemond et al.(2002)]{dullemond02}
Dullemond, C.P., van Zadelhoff, G.J. \& Natta, A.
2002, \aap, 389, 464

\bibitem[Fuente et al.(1999)]{fuente99}
Fuente, A., Mart\'in-Pintado, J., Rodr\'iguez-Fern\'andez, N.J., 
Rodr\'iguez-Franco, A., De Vicente, P., \& Kunze, D.
1999, \apj, 518, L45

\bibitem[Glassgold et al.(2004)]{glassgold04}
Glassgold, A.E., Najita, J. \& Igea, J.
2004, \apj, 615, 972

\bibitem[Glassgold et al.(2007)]{glassgold07}
Glassgold, A.E., Najita, J. \& Igea, J.
2007, \apj, 656, 515

\bibitem[Gorti \& Hollenbach(2004)]{gorti04}
Gorti, U. \& Hollenbach, D.
2004, \apj, 613, 424

\bibitem[Hern\'andez et al.(2004)]{hernandez04}
Hern\'andez, J., Calvet, N., Brice\~no, C. Hartmann, L., \& Berlind, P.
2004, \aj, 127, 1682

\bibitem[Johns-Krull et al.(2000)]{krull00}
Johns-Krull, C.M., Valenti, J.A., \& Linsky, J.L.
2000, \apj, 539, 815

\bibitem[Jonkheid et al.(2007)]{jonkheid07}
Jonkheid, B., Dullemond, C.P., Hogerheijde,M.R., \& van Dishoeck, E.F.
2007, \aap, 463, 203

\bibitem[Lacy et al.(2002)]{lacy02}
Lacy, J.H., Richter, M.J., Greathouse, T.K., Jaffe, D.T., \& Zhu, Q.
2002, \apj, 114, 153

\bibitem[Mannings \& Sargent(1997)]{mannings97}
Mannings, V., \& Sargent, A.I.
1997, \apj, 490, 792

\bibitem[Meeus et al.(2001)]{meeus01}
Meeus, G., Waters, L.B.F.M., Bouwman, J., van den Ancker, M.E., Waelkens, C., 
Malfait, K.
2001, \aap, 365, 476

\bibitem[Najita et al.(2003)]{najita03}
Najita, J., Carr, J.S., \& Mathieu, R.D. 
2003, \apj, 589, 931

\bibitem[Najita et al.(2006)]{najita06}
Najita, J.R., Carr, J.S., Glassgold, A.E.,Valenti, J.A. 2006,
in \textit{Protostars and Planets V}, ed. B. Reipurth, D. Jewitt, 
K. Keil. (Tucson: University of Arizona Press), 507

\bibitem[Nomura \& Millar(2005)]{nomura05}
Nomura, H., \& Millar, T.J.
2005, \aap, 438, 923

\bibitem[Pascucci et al.(2006)]{pascucci06}
Pascucci, I. et al. 
2006, \apj, 651, 1177

\bibitem[Qi et al.(2006)]{qi06}
Qi, C., Wilner, D.J., Calvet, N., Bourke, T.L., Blake, G.A., 
Hogerheijde, M.R., Ho, P.T.P., \& Bergin, E.
2006, \apj, 636, L157

\bibitem[Richter et al.(2002)]{richter02}
Richter, M. J., Jaffe, D. T., Blake, G. A.,
\& Lacy, J. H.
2002, \apj, 572, L161

\bibitem[Roberge et al.(2001)]{roberge01}
Roberge, A. et al.
2001, \apj, 551, L97

\bibitem[Semenov et al.(2005)]{semenov05}
Semenov, D., Pavlyuchenkov, Y., Schreyer, K., 
Henning, T., Dullemond, C., \& Bacmann, A. 
2005, \apj, 621, 853

\bibitem[Sheret et al.(2003)]{sheret03}
Sheret, I., Ramsey Howat, S.K., \& Dent, W.R.F.
2003, \mnras, 343, L65

\bibitem[Thi et al.(2001)]{thi01}
Thi, W.F. et al.
2001, \apj, 561, 1074

\bibitem[van den Ancker et al.(1998)]{vanden98}
van den Ancker, M.E., de Winter, D., \& Tjin A Djie, H.R.E.
1998, \aap, 330, 145

\end{thebibliography}
\end{document}